\documentclass[aps,pra,reprint,superscriptaddress,showpacs]{revtex4-1}

\usepackage{graphicx}
\usepackage{amssymb,amsmath}
\usepackage{bm}% bold math
\usepackage{dcolumn}

\begin{document}

\title{On the change of density of states in two-body interactions}

\author{Bo Gao}
\email{bo.gao@utoledo.edu}
\affiliation{Department of Physics and Astronomy, Mailstop 111, University of Toledo, Toledo, Ohio 43606, USA}
%\affiliation{Collaborative Innovation Center of Quantum Matter, Beijing, China}
%\affiliation{State Key Laboratory of Low-Dimensional Quantum Physics, Department of Physics, Tsinghua University, Beijing 100084, China}

\date{\today}

\begin{abstract}

We derive a general relation in two-body scattering theory that more directly relates the change of density of states (DDOS) due to interaction to the shape of the potential. The relation allows us to infer certain global properties of the DDOS from the global properties of the potential. In particular,  we show that DDOS is negative at all energies and for all partial waves, for potentials that are more repulsive than $+1/r^2$ everywhere. This behavior represents a different class of global properties of DDOS from that described by the Levinson's theorem.

\end{abstract}


\maketitle

\section{Introduction}

The density of states (DOS), or the closely related concept of the change (Delta) of density of states (DDOS) due to interaction, can be regarded as a generalization of the concept of bound spectrum to include the continuum states. It describes the energy landscape of the Hilbert space, and is one of the most important physical quantity for a quantum system. Once it is known, the equilibrium thermodynamics is completely determined, be it for a two-body, few-body, or a many-body system. The DDOS for a two-body or a few-body system also comes into play in the understanding of a many-body system through the virial expansion \cite{KU38,Huang87}, a subject that has seen a resurgence of interest in connection with the description of a unitary Fermi gas (see, e.g., Refs.~\cite{Ho04,Liu09}).

For a two-body system interacting via a central potential, the DDOS for the relative motion can be written, for each partial wave, as
\begin{equation}
\Delta D_l(\epsilon) = 
\left\{
\begin{array}{ll}
\displaystyle \sum_n \delta(\epsilon-\epsilon_{nl}) \;, & \epsilon<0 \\
\displaystyle \frac{1}{\pi}\frac{d\delta_l}{d\epsilon} \;, &\epsilon>0 
\end{array}
\;.
\right.
\label{eq:DDldef}
\end{equation}
Here $\epsilon$ is the energy in the center-of-mass frame with $\epsilon=0$ setting at the two-body threshold, $\{\epsilon_{nl}\}$ is the two-body bound spectrum for partial wave $l$, if it exists, and $\delta_l$ is the scattering phase shift for partial wave $l$. There is an additional term of $\tfrac{1}{2}\delta(\epsilon)$ for the $s$ wave in the special case of having a quasibound state right at the threshold. The $\Delta D_l$ for $\epsilon<0$ is simply a mathematical representation of its definition. Its expression in terms of phase shift for $\epsilon>0$ is due to Beth and Uhlenbeck \cite{BU37} (see also \cite{Huang87}). For continuum states with $\epsilon>0$, $\Delta D_l$ is closely related to the time-delay due to interaction, by $\Delta t=\Delta D_l/h$ \cite{Wigner55,Fano86}, where $h$ is the Planck constant. 

In terms of $\Delta D_l(\epsilon)$, the second virial coefficient $B_2$ for a single-component Bose or Fermi gas, more specifically the change of $B_2$ due to interaction, can simply be written as \cite{BU37,Huang87}
\begin{align}
\Delta B_2 &:= B_2-B_2^{(0)} \nonumber\\
	&=-2^{3/2}\lambda_T^3{\sum_l}^\prime (2l+1)\int_{-\infty}^{+\infty}d\epsilon \: \Delta D_l(\epsilon)e^{-\epsilon/k_B T} \;.
\label{eq:DB2}
\end{align}
Here $\lambda_T := (2\pi\hbar^2/mk_B T)^{1/2}$, with $k_B$ being the Boltzmann constant, is the thermal wave length of a particle with mass $m$ at temperature $T$. $B_2^{(0)}=\mp 2^{-5/2}\lambda_T^3$ are the second virial coefficients for the free (non-interacting) Bose and Fermi gases, respectively. And the ``prime'' over the summation refers to proper symmetry considerations. 

The $\Delta D_l=(1/\pi)d\delta_l/d\epsilon$ for $\epsilon>0$ gives one example that in applications of two-body scattering theory, we are often interested not only in the phase shift itself, but also in how it depends on energy. Computationally, $\Delta D_l$ for $\epsilon>0$, like most other scattering observables, is most conveniently calculated through single-channel $K$ matrix $\tan\delta_l$, in terms of which 
\begin{equation}
\Delta D_l(\epsilon>0) = \left(\frac{\mu}{\pi\hbar^2 k}\right)
	\frac{1}{1+\tan^2\delta_l}\frac{d\tan\delta_l}{dk} \;,
\label{eq:DlKl}
\end{equation}
where $\mu$ is the reduced mass, and $k$ is the length of the wave vector that is related to the energy by $\epsilon=\hbar^2k^2/2\mu$.

Instead of focusing solely on $\tan\delta_l$, as we usually do in scattering theory, our attention in this work is squarely on $d\tan\delta_l/dk$. Specifically, we derive, in Sec.~\ref{sec:theory}, a general relation relating $d\tan\delta_l/dk$ to the shape of the potential. From this relation, we show that $d\tan\delta_l/dk<0$, and therefore $d\delta_l/dk<0$ and $d\delta_l/d\epsilon<0$, for all partial waves and at all energies for any potential that is more repulsive than $+1/r^2$ for all $r$.  It means for such potentials that the phase shift is a monotonically decreasing function of energy for all partial waves. It also means for such potentials that $\Delta D_l(\epsilon)<0$ at all energies and for all partial waves. As we will discuss in Sec.~\ref{sec:implications}, this represents a different type of global property of DDOS from that implied by the Levinson's theorem \cite{Levinson49,Newton82,Ma06}. Our theoretical derivation and discussion are augmented by simple physical examples in which characteristics of DDOS are further illustrated and discussed.

\section{Theory}
\label{sec:theory}

Consider the interaction of two particles via a central potential $V(r)$. The phase shift for partial wave $l$ is determined by the solution of the radial Schr\"{o}dinger equation
\begin{equation}
\left[-\frac{\hbar^2}{2\mu}\frac{d^2}{dr^2} 
	+ \frac{\hbar^2 l(l+1)}{2\mu r^2}
	+ V(r) \right]
	u_{k l}(r) = \epsilon u_{k l}(r),
\label{eq:rsch}
\end{equation}
at a positive energy $\epsilon=\hbar^2k^2/2\mu$. This equation can be rewritten as
\begin{equation}
\left[-\frac{d^2}{dr^2} 
	+ \frac{l(l+1)}{ r^2}
	+ U(r) \right]
	u_{k l}(r) =k^2 u_{k l}(r),
\label{eq:rsch1}
\end{equation}
where $U(r) :=(2\mu/\hbar^2)V(r)$. For any potential that goes to zero as $1/r^2$ or faster at large $r$, the phase shift $\delta_l$ is determined by the solution of Eq.~(\ref{eq:rsch1}) satisfying proper physical boundary condition at the origin and with a large $r$ asymptotic behavior of 
\begin{align}
u_{kl}(r) &\sim kr\left[j_l(kr)-y_l(kr)\tan\delta_l\right]\;, \nonumber\\
	&\stackrel{r\to\infty}{\sim} \sin(kr-l\pi/2)+\cos(kr-l\pi/2)\tan\delta_l \;,
\label{eq:norm1}
\end{align}
where $j_l(x)$ and $y_l(x)$ are the spherical Bessel functions \cite{Olver10}. For any real physical system, the boundary condition at the origin is
\begin{equation}
u_{kl}(r) \stackrel{r\to 0}{\sim} 0 \;,
\label{eq:bc01}
\end{equation}
and the potential is finite everywhere with possible exception of the origin where it has to behave in a way that is consistent with Eq.~(\ref{eq:bc01}). We focus here on such physical systems, with a discussion of the model (non-physical) hard sphere potential in Sec.~\ref{sec:hs}. We would like to find out how the wave function, in particular the $\tan\delta_l$, depends on the energy, or equivalently on $k$. We are especially interested in the quantity $d\tan\delta_l/dk$, to which $\Delta D_l$ is related through Eq.~(\ref{eq:DlKl}).

Defining $z:=kr$, and dividing both sides of Eq.~(\ref{eq:rsch1}) by $k^2$, Eq.~(\ref{eq:rsch1}) and its boundary conditions, Eqs.~(\ref{eq:norm1}) and (\ref{eq:bc01}), can be written as
\begin{equation}
\left[-\frac{d^2}{dz^2} 
	+ \frac{l(l+1)}{ z^2}
	+ \frac{1}{k^2}U\left(\frac{1}{k}z\right) \right]
	u_{kl}(z) = u_{kl}(z),
\label{eq:rsch2}
\end{equation}
with
\begin{equation}
u_{kl}(z) \stackrel{z\to 0}{\sim} 0 \;.
\label{eq:bc02}
\end{equation}
and the large $z$ asymptotic behavior of
\begin{align}
u_{kl}(z) &\sim z\left[j_l(z)-y_l(z)\tan\delta_l\right]\;, \nonumber\\
	&\stackrel{z\to\infty}{\sim} \sin(z-l\pi/2)+\cos(z-l\pi/2)\tan\delta_l \;.
\label{eq:norm2}
\end{align}
This seemingly trivial rewrite, contains in fact a nontrivial alternative view on how a wave function, in particular the phase shift, depends on the energy. In the view of Eqs.~(\ref{eq:rsch1})-(\ref{eq:bc01}), the wave function is a function of $r$, the potential is fixed and the energy dependence of the phase shift are due both to the eigenvalue at the right-hand-side of Eq.~(\ref{eq:rsch1}), and to the $k$ dependence that enters into the boundary condition of Eq.~(\ref{eq:norm1}). 

In Eqs.~(\ref{eq:rsch2})-(\ref{eq:norm2}), the wave function is viewed as a function of $z$. In this view, the differential equation, Eq.~(\ref{eq:rsch2}), depends on $k$ only through the effective potential
\begin{equation}
U_\mathrm{eff}(k,z) := \frac{1}{k^2}U\left(\frac{1}{k}z\right) \;.
\end{equation}
The boundary conditions, Eqs.~(\ref{eq:bc02}) and (\ref{eq:norm2}), depend on $k$ only through the phase shift. Thus the energy dependence of the phase shift comes {\em solely} from the energy dependence of the effective potential $U_\mathrm{eff}(k,z)$.

Since the $k$ dependence of $U(z/k)$ in the $U_\mathrm{eff}$ is determined by the property of $U(r)$, and therefore $V(r)$, under a global scale transformation of $U(z) \Rightarrow U(z/k)$, it can be stated that the energy dependence of the phase shift, and therefore the DDOS due to interaction, is determined entirely by the properties of the potential under global scale transformations. This is a very general statement and conclusion, broader in scope than the relation that we are about to derive.

Now consider a solution $u_{k'l}$ at a different energy corresponding to $k'$. It satisfies
\begin{equation}
\left[-\frac{d^2}{dz^2} 
	+ \frac{l(l+1)}{ z^2}
	+ U_\mathrm{eff}(k',z) \right]
	u_{k'l}(z) = u_{k'l}(z),
\label{eq:rsch3}
\end{equation}
The Wronkian between $u_{kl}$ and $u_{k'l}$, defined by
\begin{equation}
W(u_{kl},u_{k'l}) := u_{kl}\frac{d}{dz}u_{k'l}-u_{k'l}\frac{d}{dz}u_{kl} \;,
\end{equation}
satisfies
\begin{equation}
\frac{dW}{dz} = u_{k'l}(z) \left[U_\mathrm{eff}(k',z)-U_\mathrm{eff}(k,z)\right]u_{kl}(z) \;,
\label{eq:dWdz}
\end{equation}
as a direct consequence of Eqs.~(\ref{eq:rsch2}) and (\ref{eq:rsch3}). For any physical potential, the wave functions and their derivatives, and therefore $W$, are all continuous functions of $z$ for all $z$. Integrating both sides of this equation from $z=0$ to $z=\infty$ and making use of the large $z$ asymptotic behaviors and the boundary conditions at the origin, we obtain
\begin{multline}
\tan\delta_l(k')-\tan\delta_l(k) \\
	= -\int_0^{\infty} u_{k'l}(z) \left[U_\mathrm{eff}(k',z)-U_\mathrm{eff}(k,z)\right]u_{kl}(z) dz \;.
\label{eq:DtDk}
\end{multline}
The limit of $k'\to k$ gives
\begin{equation}
\frac{d\tan\delta_l}{dk} = 
	-\int_0^{\infty} \left[u_{kl}(z)\right]^2 \left[\frac{\partial U_\mathrm{eff}(k,z)}{\partial k}\right] dz \;.
\label{eq:dtdk1}
\end{equation}
Taking the partial derivative and rewrite the equation back in terms of $V(r)$ and $r$, we have
\begin{equation}
\frac{d\tan\delta_l}{dk} =\left(\frac{2\mu}{\hbar^2}\right)\frac{1}{k^2}
	\int_0^{\infty} \left[\left(r\frac{d}{dr}+2\right)V(r)\right]\left[u_{kl}(r)\right]^2 dr \;,
\label{eq:dtdk}
\end{equation}
where the radial wave function $u_{kl}(r)$ is normalized according to Eq.~(\ref{eq:norm1}).

Equation~(\ref{eq:dtdk}) is the main mathematical result of this work. It relates $d\tan\delta_l/dk$ to a kind of ``expectation'' value of a quantity, $(r\frac{d}{dr}+2)V(r)$, which, as we will explain the next section, is a characterization of the shape of a potential \footnote{The integral in Eq.~(\ref{eq:dtdk}) is well defined (converges) for all potentials with a well defined phase shift. It is, however, not a standard expectation value, since the continuum wave function is not normalized to 1. In fact $\int_0^\infty \left[u_{kl}(r)\right]^2 dr$ diverges by itself.}. This relation allows us to arrive at certain general conclusions about the global structure of DDOS without explicitly knowledge of $\tan\delta_l$. 

\section{Implications, examples, and special cases}

\subsection{General implications}
\label{sec:implications}

The quantity $(r\frac{d}{dr}+2)V(r)$ in Eq.~(\ref{eq:dtdk}) is a characterization of the shape of a potential, more precisely a characterization of the shape in comparison to the $\pm 1/r^2$ potential. The $r\frac{d}{dr}$ operator is the generator for the global scale transformation, as in
\begin{equation}
\exp\left(\ln(\lambda) r\frac{d}{dr}\right)V(r) = V(\lambda r) \;.
\end{equation}
For a potential that is a homogeneous function of $r$, specifically $\pm C_n/r^n$ ($C_n>0$), $V(r)$ is an eigenfunction of $r\frac{d}{dr}$ with an eigenvalue of $-n$. For such potentials $(r\frac{d}{dr}+2)V(r) = \mp (n-2)C_n/r^n$, with a behavior that depends on how $n$ compares to 2. 

For the scale-invariant $\pm C_2/r^2$ potentials, $(r\frac{d}{dr}+2)V(r) \equiv 0$, and Eq.~(\ref{eq:dtdk}) shows that the corresponding phase shifts are energy-independent.

For more general potentials that are not homogeneous functions, one can write
\begin{multline}
\left(r\frac{d}{dr}+2\right)V(r) = rV(r)\left(\frac{1}{V}\frac{dV}{dr}+\frac{2}{r}\right) \\
	= rV(r)\left[\frac{1}{V}\frac{dV}{dr}-\frac{1}{(\pm C_2/r^2)}\frac{d(\pm C_2/r^2)}{dr}\right] \;.
\label{eq:comp}
\end{multline}
It again has a behavior that depends on the shape of the potential in comparison to $\pm 1/r^2$ (note that the specific value of $C_2$ does not matter). Specifically, it depends on how the log-derivative of $V$ compared to that of $\pm 1/r^2$. For a purely repulsive potential, which we define as a potential satisfying $V\ge 0$ and $dV/dr\leq 0$ (but not identically zero) for all $r$ \cite{Repulsive}, $(r\tfrac{d}{dr}+2)V(r)\leq 0$ for all $r$ describes a class of  potentials that are more repulsive than $+1/r^2$ everywhere.  Equation~(\ref{eq:dtdk}) shows for such a class of potentials that $d\tan\delta_l/dk<0$ for all energies and partial waves. It implies for such potentials that (a) the phase shifts are monotonically decreasing functions of energy for all energies and all partial waves, since $d\delta_l/d\epsilon = [\mu/\hbar^2 k (1+\tan^2\delta_l)](d\tan\delta_l/dk)$, and (b) from Eq.~(\ref{eq:DlKl}), the DDOS due to interaction is negative for all energies and all partial waves. The potentials $+C_n/r^n$ with $n>2$, to be discussed further in Sec.~\ref{sec:rn}, constitute an important subclass that falls into this category.

The behavior of $\Delta D_l<0$ for all energies, namely DOS being uniformly reduced, represents a different type of global property of the energy landscape of the Hilbert space from that implied by the Levinson's theorem \cite{Levinson49,Newton82,Ma06}. For potentials satisfying $\int_0^\infty dr r|V(r)| <\infty$, which we will call the Newton criterion \cite{Newton82,Ma06}, the phase shift satisfy
\[
\delta_l(0)-\delta_l(\infty) = n_l\pi \;,
\]
where $n_l$ is the number of bound states for partial wave $l$ ($n_l$ is replaced by $n_l+1/2$ for the special case of $s$ wave with a quasibound state right at the threshold). In terms of the DDOS, Levinson's theorem implies
\begin{equation}
\int_{-\infty}^{+\infty} \Delta D_l d\epsilon = 0 \;.
\end{equation}
Thus for all potentials satisfying the Newton criterion, the density of states is \textit{redistributed}. If DOS is reduced by the interaction over one range of energies, it has to be enhanced over other energies, in a way that preserves the total number of states (see the example of Sec.~\ref{sec:rsqw}). This redistribution of DOS by the interaction generally leads to a complex energy landscape in the Hilbert space, and is a major source of the complexity of an interacting quantum system. The potentials that are everywhere more repulsive than $+1/r^2$ can violate the Levinson's theorem because they do not satisfy the Newton criterion.

\subsection{The example of $+C_n/r^n$ potentials with $n>2$}
\label{sec:rn}

Among homogeneous potentials, $\pm C_n/r^n$, only those with $n\ge 2$ have phase shift as defined by Eq.~(\ref{eq:norm1}), and among this subset, the ones of most physical interest are those with an integer $n$, especially those with $n=3,4$, and 6. 

%\subsection{The example of a repulsive $1/r^3$ potential}

\begin{figure}
\includegraphics[width=\columnwidth]{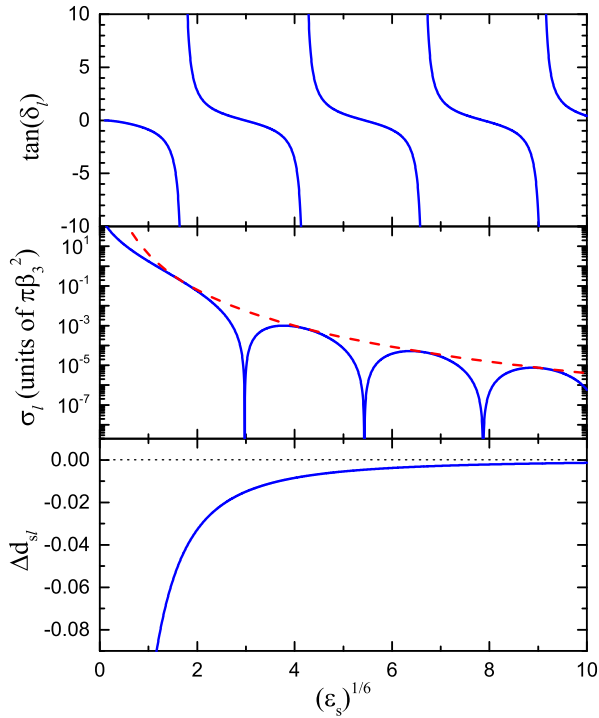}
\caption{(Color online) $s$ wave scattering properties for a repulsive $+C_3/r^3$ potential, as a function of a scaled energy. Here $\beta_3:=2\mu C_3/\hbar^2$ is the length scale associated with the potential, $s_E := (\hbar^2/2\mu)(1/\beta_3)^2$ is the corresponding energy scale, and $\epsilon_s:=\epsilon/s_E$ is the scaled energy. (a) Single-channel $K$ matrix $\tan\delta_l$. (b) Partial cross section $\sigma_l$ (solid line) and its corresponding unitarity limit (dashed line) given by $(2l+1)4\pi/k^2$. (c) Change of the density of states due to interaction as represented by the scaled $\Delta d_{sl}:=\Delta d_l/\beta_3$. 
\label{fig:purer3s}}
\end{figure}

For a repulsive homogeneous potential, $V(r)=+C_n/r^n$ ($C_n>0$), Eq.~(\ref{eq:dtdk}) gives
\begin{equation}
\frac{d\tan\delta_l}{dk} =-(n-2)\left(\frac{2\mu C_n}{\hbar^2}\right)\frac{1}{k^2}\int_0^{\infty} \frac{1}{r^n}\left[u_{kl}(r)\right]^2 dr \;.
\label{eq:dtdkrn}
\end{equation}
It shows more explicitly that for all repulsive homogeneous potentials with $n>2$, $d\tan\delta_l/dk<0$ and therefore $\Delta D_l<0$, for all energies and all partial wave, as stated earlier in a more general context. 

Figure~\ref{fig:purer3s} illustrates some of the $s$ wave ($l=0$) scattering properties for a repulsive $+C_3/r^3$ potential \cite{Gao99a}. The $\tan\delta_l$, shown in Fig.~\ref{fig:purer3s}(a), is evaluated from Eq.~(44) of Ref.~\cite{Gao99a}, from which the partial cross section, shown in Fig.~\ref{fig:purer3s}(b), is evaluated from 
\begin{equation}
\sigma_l = \frac{4\pi}{k^2}(2l+1)\frac{\tan^2\delta_l}{1+\tan^2\delta_l} \;.
\label{eq:pxs}
\end{equation}
Instead of $\Delta D_l$, Fig.~\ref{fig:purer3s}(c) shows the closely related $\Delta d_l$, the change of the number of states per unit $k$. They are related by
\begin{align}
\Delta d_l &:= \frac{1}{\pi}\frac{d\delta_l}{dk} \nonumber\\
& = \frac{1}{\pi}\frac{1}{1+\tan^2\delta_l}\frac{d\tan\delta_l}{dk} \nonumber\\
& = \left(\frac{\hbar^2k}{\mu}\right)\Delta D_l \;.
\end{align}
In evaluating $\Delta d_l$,  $d\tan\delta_l/dk$ is again evaluated from Eq.~(44) of Ref.~\cite{Gao99a}.

Figure~\ref{fig:purer3s}(c) shows that $\Delta d_l$, and therefore $\Delta D_l$, is negative for all energies, as expected. It also implies that the phase shift is a monotonically decreasing function of energy, a fact that is also embedded in Figure~\ref{fig:purer3s}(a). Figure~\ref{fig:purer3s}(b) shows that such a simple behavior of phase shift is not obvious if one looks only at the partial cross section. As we will also see from other examples, the DDOS has generally a simpler behavior than the cross section and gives a better picture of the structure of the continuum states. Even for a monotonically decreasing phase shift, the partial cross section can have structures associated with diffraction resonances \cite{Gao10a,Gao13c,LYG14}.

Both $\Delta D_l$ and $\Delta d_l$ have of course the same physical content. The $\Delta D_l$ has a cleaner physical interpretation, especially in connection with the bound spectrum [see, Eq.~(\ref{eq:DDldef})]. If the focus is solely on the continuum states, e.g. for repulsive potentials which have no bound states, the quantity $\Delta d_l$ can be more convenient for mathematical and illustration purposes. For any potential that follows the Wigner threshold behavior \cite{Wigner48} for the $s$ wave, as described by the effective range theory \cite{Schwinger47,Blatt49,Bethe49}, $\tan\delta_{l=0}\sim -a_{l=0}k$ in the limit of $k\to 0$, where $a_{l=0}$ is the $s$ wave scattering length. Thus $\Delta D_{l=0}$ has, for all such potentials, a $1/k$ singularity at zero energy for the $s$ wave, while $\Delta d_{l=0}$ is well behaved and goes to a constant (see the next two examples in the following subsections). Specifically, $\Delta d_{l=0}\sim -a_{l=0}/\pi$. This behavior is followed by all potentials that goes to zero faster than $1/r^3$ at large $r$ \cite{LK63}. The example of $+C_3/r^3$ gives an exception. In Fig.~\ref{fig:purer3s}(c), a singularity at zero energy remains even in $\Delta d_{l=0}$, an indication of the breakdown of the Wigner threshold behavior. Indeed for a $+C_3/r^3$ potential, the Wigner threshold behavior is violated, and we have in particular $\tan\delta_{l=0}\sim (k\beta_3)\ln(k\beta_3)$, where $\beta_3:=2\mu C_3/\hbar^2$ is the length scale associated with the $+C_3/r^3$ potential \cite{Gao99a}. 

This example also helps to make the point that the main application of our mathematical result, Eq.~(\ref{eq:dtdk}), is to understand the global properties of $d\tan\delta_l/dk$, not necessarily for its specific value at a particular energy. The evaluation of the RHS of Eq.~(\ref{eq:dtdk}) requires the knowledge of the wave function over an extended range of $r$, which is generally much more expensive than finding $\tan\delta_l$, which requires only the large $r$ asymptotic behavior of the wave function. Computationally, the value of $d\tan\delta_l/dk$ is still more easily obtained by computing $\tan\delta_l$ as a function of energy as in standard scattering theory. 

\subsection{The example of a repulsive finite square well potential}
\label{sec:rsqw}

\begin{figure}
\includegraphics[width=\columnwidth]{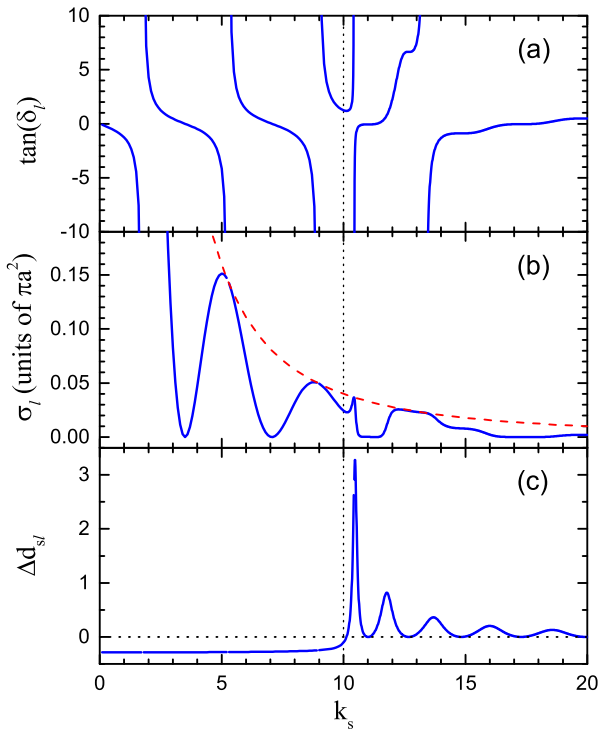}
\caption{(Color online) $s$ wave scattering properties for a repulsive finite square well with $k_{0s} := k_0 a = 10.0$, as a function of $k_s :=ka$. The location of $k_s=k_{0s}$, represented by a vertical dotted line, corresponds to the energy being equal to the height of the potential $V_0$. (a) Single-channel $K$ matrix $\tan\delta_l$. (b) Partial cross section (solid line) and the corresponding unitarity limit (dashed line) given by $(2l+1)4\pi/k^2$. (c) Change of the density of states due to interaction as represented by the scaled $\Delta d_{sl}:= \Delta d_{l}/a$. 
\label{fig:rsqw10s}}
\end{figure}

We use the familiar example of a finite square well potential, for which everything can be worked out in simple analytical forms, to further illustrate a number of characteristics of DDOS due to interaction. In particular, we use this example to show that a repulsive potential \cite{Repulsive} does not automatically guarantee a reduction of DOS for all energies. The example also give an illustration of how the Levinson's theorem manifests itself.

A repulsive finite square well potential, defined by
\begin{equation}
V(r) = \left\{
\begin{array}{ll}
V_0 \;, & r<a \\
0 \;, & r>a
\end{array}
\right.\;,
\end{equation}
with $V_0>0$ and $a>0$, is a purely repulsive potential in the sense of $V\ge 0$ and $dV/dr\leq 0$ for all $r$ \cite{Repulsive}. For this potential, both the wave function $u_{kl}$ and the $\tan\delta_l$ are well known (see, e.g., Ref.~\cite{Joachain75}). In particular, the $\tan\delta_l$ can be written as
\begin{equation}
\tan\delta_l = \frac{k_s j_l^\prime(k_s)-\gamma_l(\bar{k}_s) j_l(k_s)}
	{k_s y_l^\prime(k_s)-\gamma_l(\bar{k}_s) y_l(k_s)} \;,
\end{equation}
where $k_s:=ka$ and
\[
\gamma_l(\bar{k}_s) := \bar{k}_s j_l^\prime(\bar{k}_s)/j_l(\bar{k}_s) \;,
\]
in which $\bar{k}_s:=[k_s^2-(k_0a)^2]^{1/2}$ and we have defined $k_0:=(2\mu V_0/\hbar^2)^{1/2}$. This result is applicable both for energies above the height of the potential $V_0$ where $k_s>k_{0s}:=k_0 a$, and for energies below $V_0$ where $k_s<k_{0s}$ and $\bar{k}_s=i|\bar{k}_s|$.  Except for the low-energy and the high-energy limits, where the results are used to check the effective-range theory \cite{Schwinger47,Blatt49,Bethe49} and the Born approximation respectively (see, e.g., Ref.~\cite{Joachain75}), the interesting physics embedded in this analytic solution in the intermediate energy regime does not seem to have been much discussed.

Figure~\ref{fig:rsqw10s} illustrates some of the physical quantities of interest for the $s$ wave scattering by a repulsive finite square well potential. After a simple scaling based on the length scale $a$, the potential is characterized by a single strength parameter $k_{0s}:=k_0 a$, which for our example is taken to be $k_{0s}=10.0$. Figure~\ref{fig:rsqw10s}(a) illustrates the $\tan\delta_l$. Figure~\ref{fig:rsqw10s}(b) illustrates the partial cross section given by Eq.~(\ref{eq:pxs}). For this particular example, the integral in Eq.~(\ref{eq:dtdk}) can be easily carried out. The $d\tan\delta_l/dk$ can be obtained either from Eq.~(\ref{eq:dtdk}) using the known wave function or directly from the known $\tan\delta_l$. They have been used to check for consistency against each other. From $d\tan\delta_l/dk$, we obtain for the DDOS, specifically the scaled $\Delta d_{sl}:= \Delta d_{l}/a$,
\begin{widetext}
\begin{equation}
\Delta d_{sl} = -\frac{k_{0s}^2}{\pi k_s^2(k_s^2-k_{0s}^2)}
	\frac{l(l+1)-\gamma_l(\bar{k}_s)[\gamma_l(\bar{k}_s)+1]}
	{[k_s j_l^\prime(k_s)-\gamma_l(\bar{k}_s) j_l(k_s)]^2
	+[k_s y_l^\prime(k_s)-\gamma_l(\bar{k}_s) y_l(k_s)]^2} \;,
\end{equation}
\end{widetext}
which is illustrated in Figure~\ref{fig:rsqw10s}(c) for the $s$ wave.

Recall that the DDOS for a continuum state is related directly to the time delay (advance) in scattering by $\Delta t=\Delta D_l/h$ \cite{Wigner55,Fano86}. Thus an increase in DOS corresponds to a time delay due to interaction, while a decrease in DOS corresponds to a time advance. If one were to combine this interpretation with classical mechanics, one would arrive at a seemingly reasonable conclusion that all purely repulsive potentials \cite{Repulsive} should lead to time advance and therefore reduction of DOS for all energies and all partial waves (impact parameters). If this were true, our earlier conclusion would seem be trivial with little real physical content.

Figure~\ref{fig:rsqw10s}(c) shows that this classical expectation is incorrect. While DOS is indeed reduced for $\epsilon<V_0$, for most energies above $V_0$, it is enhanced by the interaction. This gives an example that even for a purely repulsive potential there can generally be regions of energies where DOS is enhanced if the condition of $(r\tfrac{d}{dr}+2)V(r)<0$ is not satisfied for all $r$. This result is a manifestation of the Levinson's theorem \cite{Levinson49,Newton82,Ma06}. A repulsive finite square well satisfies the Newton criteria \cite{Newton82,Ma06} and therefore the Levinson's theorem. For all such potentials, if the DOS is reduced over one range of energies, in this case below $V_0$, it has to be enhanced over other ranges of energies, in this case above $V_0$. From a different angle, the quantum effects come specifically in this case from the quantum reflection at the boundary $a$. It is the quantum reflection and the resulting interference pattern \cite{Gao08a} that gives rise to the resonances seen in Figure~\ref{fig:rsqw10s} above $V_0$. Whatever we choose to call those resonances, it should be clear that they are fundamentally of the same physical origin as shape resonances  \cite{Gao08a,Gao13c}, even though there is no visible potential barrier in this case. Below $V_0$, where the phase shift is a monotonically decreasing function of energy, there is again structure in the cross section that is associated with diffraction resonances \cite{Gao10a,Gao13c,LYG14}. Last but not the least, we note that as a result of the redistribution of DOS, the energy landscape of the Hilbert space, which is best described by DDOS, has become much more complex compared to the earlier example where the DOS is uniformly reduced.

\subsection{The special case of a hard sphere potential}
\label{sec:hs}

\begin{figure}
\includegraphics[width=\columnwidth]{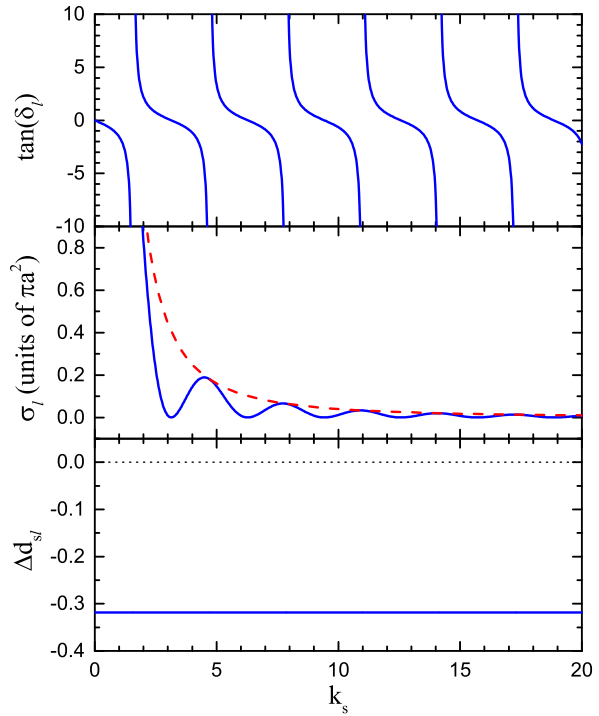}
\caption{(Color online) $s$ wave scattering properties for a hard sphere potential, as a function of $k_s:=ka$. (a) Single-channel $K$ matrix $\tan\delta_l$. (b) Partial cross section (solid line) and the corresponding unitarity limit (dashed line) given by $(2l+1)4\pi/k^2$. (c) Change of the density of states due to interaction as represented by the scaled $\Delta d_{sl}:= \Delta d_{l}/a$. 
\label{fig:hsl0}}
\end{figure}

The hard sphere potential
\begin{equation}
V(r) = \left\{
\begin{array}{ll}
\infty \;, & r\leq a \\
0 \;, & r>a
\end{array}
\right.\;,
\label{eq:Vhs}
\end{equation}
with $a>0$ is not a physical potential because it is infinite for $r\leq a$. It is nevertheless a very useful model potential especial for simplifying the description of interaction in a few-body or a many-body quantum system. For the hard sphere potential, Eq.~(\ref{eq:dtdk}) is not applicable. A modified version is required, but is easily derived along the same approach of Sec.~\ref{sec:theory}. 

For $r>a$, $V\equiv 0$, and the dependences of the wave function and the phase shift on the energy are due solely to the energy dependence of the boundary condition. Specifically, the boundary condition, $u_{kl}(r=a)=0$, when viewed through $u_{kl}(z)$, is $u_{kl}(z=ka) = 0$, which is energy dependent. Integrating Eq.~(\ref{eq:dWdz}) from $z' = k'a>ka$ to $z=\infty$ now gives us
\begin{equation}
\frac{d\tan\delta_l}{dk} = -\frac{a}{k^2}\left(\left.\frac{du_{kl}}{dr}\right|_{r=a}\right)^2 \;.
\label{eq:dtdkhs}
\end{equation}
It shows for a hard sphere potential that $d\tan\delta_l/dk<0$ and therefore $\Delta D_l<0$ for all energies and all partial waves, without any explicit calculation. In this global behavior, a hard sphere belongs, not surprisingly, to the same class as other repulsive potentials that are more repulsive than $+1/r^2$. 

Equation~(\ref{eq:dtdkhs}) is easily verified using the known wave function and phase shifts for a hard sphere [keeping in mind that the wave function is normalized according to Eq.~(\ref{eq:norm1})]. Figure~\ref{fig:hsl0} illustrates some of its $s$ wave scattering properties. From the well known (see, e.g., Ref.~\cite{Joachain75})
\begin{equation}
\tan\delta_l = \frac{j_l(k_s)}{y_l(k_s)} \;,
\end{equation}
where $k_s:=ka$, the cross section is obtained from Eq.~(\ref{eq:pxs}). One also obtains, for $\Delta d_{sl}:= \Delta d_{l}/a$,
\begin{equation}
\Delta d_{sl} = -\frac{1}{\pi k_s^2[j_l^2(k_s)+y_l^2(k_s)]} \;,
\end{equation}
which is negative for all energies. Figure~\ref{fig:hsl0} again illustrates that the DDOS has generally much simpler structure than other quantities used to describe the scattering continuum, and the cross section has structure associated with diffraction resonances \cite{Gao10a,Gao13c,LYG14}, despite of the phase shift being a monotonically decreasing function of energy.

\section{Discussion and conclusions}

In conclusion, we have derived a general relation between $d\tan\delta_l/dk$ and the shape of a potential. From this relation, we show that there exists a class of physical potentials, such as  $+C_n/r^n$ with $n>2$, for which the density of states is uniformly reduced by the interaction, namely $\Delta D_l<0$ for all energies and all partial waves. This behavior represents a different kind a global property of the Hilbert space under interaction. It differs from that characterized by the Levinson's theorem \cite{Levinson49,Newton82,Ma06} which is followed by most other physical potentials, including all potentials that are attractive at large distances with a behavior of $-1/r^n$ with $n>2$.

This result suggests a broad classification of interaction potentials into two classes. For one class, such as  $+C_n/r^n$ with $n>2$, the DOS is uniformly reduced, and the energy landscape of the Hilbert space remains flat despite interaction. For the other class, which includes all potentials that satisfy Levinson's theorem, the DOS is redistributed, leading generally to a Hilbert space that has a much more complex energy landscape with peaks and valleys. 
Our conjecture, which is also the motivation behind this work, is that this qualitative relation between the global structure of energy landscape and the shape of interaction potential should remain valid much more generally in an $N$-body quantum system. Specifically, we expect that the DDOS due to interaction should be uniformly negative if interaction potentials between all particles are all more repulsive than $+1/r^2$.

This work is one of our first steps in an effort to better understand the complexities of an $N$-body interacting quantum system. In future works, we hope to also address the complexities due to attractive interactions, in both two-body and $N$-body contexts.

\begin{acknowledgments}
This work is supported by NSF under Grant Nos. PHY-1306407 and PHY-1607256. 
\end{acknowledgments}

\bibliography{bgao,qdt,twobody,virial}

\begin{thebibliography}{24}
\bibitem [{Rep()}]{Repulsive}%
  \BibitemOpen
  \href@noop {} {}\bibinfo {note} {In quantum mechanics, it has been quite
  common to call any potential with $V\ge 0$ (but not zero everywhere) as being
  repulsive (see, e.g., Ref.~\cite{Joachain75}). We believe that this is due at
  least partly to the fact that $dV/dr$ has never been shown to be relevant to
  the degree of this work for the continuum states. Our definition of a purely
  repulsive potential, $V\ge 0$ and $dV/dr\leq 0$ for all $r$, is consistent
  with that of classical mechanics. When necessary to distinguish, we will call
  the $V\ge 0$ only requirement as the weaker version of being
  repulsive.}\BibitemShut {Stop}%
\end{thebibliography}

\end{document}